\documentclass[a4paper,12pt]{article}
\usepackage{amssymb}
\usepackage{latexsym}
\usepackage[dvips]{graphicx}
\usepackage{cite}
\newcommand{\be}{\begin{equation}}
\newcommand{\ee}{\end{equation}}
\newcommand{\bea}{\begin{eqnarray}}
\newcommand{\nn}{\nonumber}
\newcommand{\eea}{\end{eqnarray}}

\begin{document}

\begin{titlepage}
\begin{flushright}
\end{flushright}
\begin{centering}
\vspace{.3in}
{\Large{\bf Energy distribution in the dyadosphere\\
of a Reissner-Nordstr\"om black hole in M{\o}ller's prescription}}
\\

\vspace{.5in} {\bf  Elias C.
Vagenas\footnote{evagenas@phys.uoa.gr} }\\

\vspace{0.3in}

Nuclear and Particle Physics Section\\
Physics Department\\
University of Athens\\
GR-15771, Athens, Greece\\

\end{centering}

\vspace{0.7in}
%%%%%%%%%%%%%%%%%%%ABSTRACT%%%%%%%%%%%%%%%%%%%%%%%%%%%%%%%%%%
\begin{abstract}
\par\noindent
The energy and momentum distributions in the dyadosphere of a
Reissner-Nordstr\"om black hole  are evaluated. The M{\o}ller's
energy-momentum complex is employed for this computation. The
spacetime under study is modified due to the effects of vacuum
fluctuations in the dyadosphere. Therefore, the corrected
Reissner-Nordstr\"om black hole metric takes
into account the first contribution of the weak field limit of
one-loop QED. Furthermore, a comparison and a consequent
connection between our results that those already existing in the
literature is provided. We hypothesize that when the energy distribution is of specific form
there is a relation that connects the coefficients
in the Einstein's prescription with those in the M{\o}ller's prescription.

\end{abstract}

%%%%%%%%%%%%%%%%%%%%%%%%%%%%%%%%%%%%%%%%
\end{titlepage}
\newpage

\baselineskip=18pt
%%%%%%%%%%%%%%%%%%%%%%%%%%%%%%%%%%%%%%%%%%%%%%%%%%%%%%%%%%%%%%%%%%%%%%%%%%%%%%%%%%%%%%%%%%%%%%%%%
%%%%%%%%%%%%%%%%%%%%%%%%%%%%%%%%%%%%%%%%%%%%%%%%%%%%%%%%%%%%%%%%%%%%%%%%%%%%%%%%%%%%%%%%%%%%%%%%
%%%%%%%%%%%%%%%%%%%%%%%%%%%%%%%%%%%%%%%%%%%%%%%%%%%%%%%%%%%%%%%%%%%%%%%%%%%%%%%%%%%%%%%%%%%%%%%%%
%%%%%%%%%%%%%%%%%%%%%%%%%%% INTRODUCTION %%%%%%%%%%%%%%%%%%%%%%%%%%%%%%%%%%%%%%%%%%%%%%%%%%%%%%%%
\section*{Introduction}
Energy-momentum localization has been one of the most interesting
but also thorny problems for the General Theory of Relativity. A
plethora of different attempts to solve this problem have led to
inconclusive results till now. Energy-momentum complexes
introduced first by Einstein, were the foremost endeavor to solve
this problem. After that a large number of different expressions
for the energy-momentum complexes were proposed. A drawback of
this attempt was that energy-momentum complexes had to be computed
in quasi-Cartesian coordinates. M{\o}ller proposed a new
expression for an energy-momentum complex which could be utilized
to any coordinate system. However, the idea of the energy-momentum
complex was severely criticized for a number of reasons.
Considerable attempts to deal with this problematic issue are also
the quasi-local and the superenergetic quantities\footnote{For a
list of references concerning all the aforementioned issues see
\cite{elias2}.}.

\par\noindent
Virbhadra and collaborators enlivened anew the concept of energy-momentum complexes \cite{virbhadra}.
Since then, numerous works have been performed on evaluating the energy and momentum distributions
of several gravitational backgrounds using the energy-momentum complexes \cite{complexes}.
In support of the importance of the concept of energy-momentum complexes,
Chang, Nester and Chen \cite{nester} proved that every energy-momentum complex is associated with a
Hamiltonian boundary term. Thus, the energy-momentum complexes are quasi-local and acceptable.

\par\noindent
In this paper we evaluate the energy and momentum density distributions of the dyadosphere of the modified
Reissner-Nordstr\"om black hole metric. The modifications to the aforesaid metric show up because of the
the first contribution of the weak field limit of one-loop QED.
The prescription that is used in the present analysis, is the one introduced by M{\o}ller.
The reasons for presenting here the M{\o}ller's description are:
(a) the argument that it is not restricted to quasi-Cartesian coordinates and (b) a
work of Lessner \cite{lessner} who argues that the M{\o}ller's energy-momentum complex is a powerful concept of energy
and momentum in General Theory of Relativity.

\par\noindent
The remainder of the paper is organized as follows.
In Section 1 we consider the concept of energy-momentum complexes in the framework of General Theory of Relativity.
In Section 2 the M{\o}ller's energy-momentum complex is presented.
In Section 3 we introduce the concept of dyadosphere for the Reissner-Nordstr\"om black hole and present the modified
Reissner-Nordstr\"om black hole metric.
In Section 4 we provide all results that already exist in the literature.
In Section 5 we utilize M{\o}ller's prescription and thus we calculate the energy and momentum density distributions
of the afore-mentioned spacetime. Furthermore, we compare order by order the coefficients that appear in the expression for the energy
in the M{\o}ller's prescription with those that appear when other prescriptions are utilized.
A relation that connects M{\o}ller's coefficients with those of other prescriptions is given.
Finally, Section 6 is devoted to a brief summary of results and concluding remarks.
%%%%%%%%%%%%%%%%%%%%%%%%%%%%%%%%%%%%%%%%%%%%%%%%%%%%%%%%%%%%%%%%%%%%%%%%%%%%%%%%%%%%%%%%%%%%%%%%%
%%%%%%%%%%%%%%%%%%%%%%%%%%%%%%%%%%%%%%%%%%%%%%%%%%%%%%%%%%%%%%%%%%%%%%%%%%%%%%%%%%%%%%%%%%%%%%%%%
%%%%%%%%%%%%%%%%%%%%%%%%%%%%%%%%%%%%%%%%%%%%%%%%%%%%%%%%%%%%%%%%%%%%%%%%%%%%%%%%%%%%%%%%%%%%%%%%%
\section{Energy-Momentum Complexes}
The conservation laws of energy and momentum  for an isolated (closed), i.e. no external force acting on the
system, physical system in the Special Theory of Relativity are expressed by a set of differential equations.
Defining $T^{\mu}_{\nu}$ as the symmetric energy-momentum tensor of matter and all non-gravitational fields the
conservation laws are given by
\be
T^{\mu}_{\nu,\, \mu} \equiv \frac{\partial T^{\mu}_{\nu}}{\partial
x^{\mu}}=0\ee where \be \rho=T^{t}_{t}\hspace{1cm}j^{i}=T^{i}_{t}\hspace{1cm}p_{i}=-T^{t}_{i}
\ee
are the energy density, the energy current density, the momentum density,
respectively, and Greek indices run over the spacetime
labels while Latin indices run over the spatial coordinate values.

\par\noindent
Making the transition from the Special to General Theory of Relativity one adopts a simplicity principle which is
called principle of minimal gravitational coupling. As a result of this, the conservation equation is now written
as
\be
T^{\mu}_{\nu;\, \mu} \equiv \frac{1}{\sqrt{-g}}\frac{\partial}{\partial
x^{\mu}}\left(\sqrt{-g}\,T^{\mu}_{\nu}\right)-\Gamma^{\kappa}_{\nu\lambda}T^{\lambda}_{\kappa}=0
\ee
where $g$ is the determinant of the metric tensor $g_{\mu\nu}(x)$. The conservation equation may also be written as
\be
\frac{\partial}{\partial x^{\mu}}\left(\sqrt{-g}\,T^{\mu}_{\nu}\right)=\xi_{\nu}
\ee
where
\be
\xi_{\nu}=\sqrt{-g}\Gamma^{\kappa}_{\nu\lambda}T^{\lambda}_{\kappa}
\ee
is a non-tensorial object. For $\nu=t$ this means that the matter energy is not a conserved quantity for
the physical system\footnote{It is possible to restore the conservation law by introducing a local
inertial system for which at a specific spacetime point $\xi_{\nu}=0$ but this equality by no means holds in general.}.
From a physical point of view this lack of energy conservation can
be understood as the possibility of transforming matter energy into gravitational energy and vice versa. However,
this remains a problem and it is widely believed that in order to be solved one has to take into account the
gravitational energy.
\par\noindent
By a well-known procedure, the non-tensorial
object $\xi_{\nu}$ can be written  as
\be
\xi_{\nu}=-\frac{\partial}{\partial
x^{\mu}}\left(\sqrt{-g}\,\vartheta^{\mu}_{\nu}\right)
\ee
where $\vartheta^{\mu}_{\nu}$ are certain functions of
the metric tensor and its first order derivatives. Therefore, the energy-momentum tensor of matter $T^{\mu}_{\nu}$
is replaced by the expression
\be
\theta^{\mu}_{\nu}=\sqrt{-g}\left(T^{\mu}_{\nu}+\vartheta^{\mu}_{\nu}\right)
\ee
which is called energy-momentum complex since it is a combination of the tensor $T^{\mu}_{\nu}$ and a pseudotensor
$\vartheta^{\mu}_{\nu}$ which describes the energy and  momentum of the gravitational field. The energy-momentum
complex satisfies a conservation law in the ordinary sense, i.e.
\be
\theta^{\mu}_{\nu,\, \mu}=0
\ee
and it can be
written as
\be
\theta^{\mu}_{\nu}=\chi^{\mu\lambda}_{\nu,\,\lambda}
\ee
where $\chi^{\mu\lambda}_{\nu}$ are called
superpotentials and are functions of the metric tensor and its first order derivatives.

\par\noindent
It is obvious that the energy-momentum complex is not
uniquely determined by the condition that is usual divergence is zero since it can always been added to the
energy-momentum complex a quantity with an identically vanishing divergence.

%%%%%%%%%%%%%%%%%%%%%%%%%%%%%%%%%%%%%%%%%%%%%%%%%%%%%%%%%%%%%%%%%%%%%%%%%%%%%%%%%%%%%%%%%%%%%%%%%
%%%%%%%%%%%%%%%%%%%%%%%%%%%%%%%%%%%%%%%%%%%%%%%%%%%%%%%%%%%%%%%%%%%%%%%%%%%%%%%%%%%%%%%%%%%%%%%%%
%%%%%%%%%%%%%%%%%%%%%%%%%%%%%%%%%%%%%%%%%%%%%%%%%%%%%%%%%%%%%%%%%%%%%%%%%%%%%%%%%%%%%%%%%%%%%%%%%
\section{M{\o}ller's Prescription}
The energy-momentum complex of M{\o}ller in a  four-dimensional background
is given as \cite{moller}
\be
\mathcal{J}^{\mu}_{\nu}=\frac{1}{8\pi}\xi^{\mu\lambda}_{\nu\,\, ,\, \lambda}
\label{mtheta}
\ee
where the M{\o}ller's superpotential $ \xi^{\mu\lambda}_{\nu}$ is of the form
\be
\xi^{\mu\lambda}_{\nu}=\sqrt{-g}
\left(\frac{\partial g_{\nu\sigma} }{\partial x^{\kappa} }-
\frac{\partial g_{\nu\kappa}}{\partial x^{\sigma}
}\right)g^{\mu\kappa}g^{\lambda\sigma}
\label{msuper}
\ee
with the antisymmetric property
\be
\xi^{\mu\lambda}_{\nu}=-\xi^{\lambda\mu}_{\nu}\hspace{1ex}.
\ee

\par\noindent
It is easily seen that the M{\o}ller's energy-momentum complex
satisfies the local conservation equation
\be
\frac{\partial \mathcal{J}^{\mu}_{\nu}}{\partial x^{\mu}}=0
\ee
where  $\mathcal{J}^{0}_{0}$ is the energy density and $\mathcal{J}^{0}_{i}$ are the momentum density components.

\par\noindent
Thus, the energy and momentum in M{\o}ller's
prescription for a four-dimensional background are given by
\be
P_{\mu}=\int\int\int
\mathcal{J}^{0}_{\mu}dx^{1}dx^{2}dx^{3}
\label{mmomentum}
\ee
and specifically the energy of the physical system in a
four-dimensional background is
\be
E=\int\int\int
\mathcal{J}^{0}_{0}dx^{1}dx^{2}dx^{3}\hspace{1ex}.
\label{menergy}
\ee

\par\noindent
It should be noted that the calculations are not anymore restricted
to quasi-Cartesian coordinates but they can be
utilized in any coordinate system.
%%%%%%%%%%%%%%%%%%%%%%%%%%%%%%%%%%%%%%%%%%%%%%%%%%%%%%%%%%%%%%%%%%%%%%%%%%%%%%%%%%%%%%%%%%%%%%%%%
%%%%%%%%%%%%%%%%%%%%%%%%%%%%%%%%%%%%%%%%%%%%%%%%%%%%%%%%%%%%%%%%%%%%%%%%%%%%%%%%%%%%%%%%%%%%%%%%%
%%%%%%%%%%%%%%%%%%%%%%%%%%%%%%%%%%%%%%%%%%%%%%%%%%%%%%%%%%%%%%%%%%%%%%%%%%%%%%%%%%%%%%%%%%%%%%%%%
%%%%%%%%%%%%%%%%%%%%%%%%%%%%%%%%%%%%%%%%%%%%%%%%%%%%%%%%%%%%%%%%%%%%%%%%%%%%%%%%%%%%%%%%%%%%%%%%%
%%%%%%%%%%%%%%%%%%%%%%%%%%%%%%%%%%%%%%%%%%%%%%%%%%%%%%%%%%%%%%%%%%%%%%%%%%%%%%%%%%%%%%%%%%%%%%%%%
%%%%%%%%%%%%%%%%%%%%%%%%%%%%%%%%%%%%%%%%%%%%%%%%%%%%%%%%%%%%%%%%%%%%%%%%%%%%%%%%%%%%%%%%%%%%%%%%%
\section{The Dyadosphere of the Reissner-Nordstr{\"o}m black hole}
The Reissner-Nordstr{\"o}m black hole is described by the line element
\be
ds^2 = f(r)dt^2 - \frac{dr^2}{f(r)} -r^2 (d\theta^2 + \sin\theta d\phi^2)
\label{metric}
\ee
where the metric element $f(r)$ is given as
\be
f(r)=1-\frac{2 M}{r} + \frac{Q^2}{r^2}
\label{metric1}
\ee
and the quantities $M$ and $Q$ are the mass and the electric charge, respectively, of the black hole.
The black hole horizon of the Reissner-Nordstr{\"o}m black hole is located at\footnote{We have set $G=1$
and $c=1$.}
\be
r_{+}=M\pm \sqrt{M^2 -Q^2}
\hspace{1ex}.
\ee
The concept of dyadosphere was first introduced by Ruffini \cite{Ruffini:1998df}, and also Preparata, Ruffini,
and Xue \cite{Preparata:1998rz}.
They claimed that outside the black hole horizon there is a region where the electromagnetic field strength, namely
the only nonvanishing radial component ${\cal \vec{E}} =\frac{Q}{ r^2}\hat{r}$,
is larger than the well-known Heisenberg-Euler critical value for the electron-positron pair creation
\be
{\cal E}_{crit} =\frac{m_{e}^{2}c^{3}}{\hbar e}
\ee
where $m_e$ and $e$ stand for the mass and the electric charge, respectively, of an electron.
This region is called dyadosphere and it extends from the black hole horizon, i.e. $r_+$, which is the inner radius
of the dyadosphere to an outer radius, $r_{ds}$, given by\footnote{Here we have reinstated the constants $G$ and $c$ in order
to make clear the hybrid gravitational and quantum nature of this quantity.}
\be
r_{ds}=\sqrt{\left(\frac{\hbar}{m_{e}c}\right)\left(\frac{GM}{c^2}\right)
\left(\frac{m_{pl}}{m_e}\right)\left(\frac{e}{q_{pl}}\right)\left(\frac{Q}{\sqrt{G}M}\right)}
\ee
where $m_{pl}$ $\left(=\sqrt{\frac{\hbar c}{G}}\right)$ and $q_{pl}$ $\left(=\sqrt{\hbar c}\right)$
are respectively the Planck mass and charge. It was found that the dyadosphere exists only for
black holes with mass from the upper limit for neutron stars at
$~ 3.2 M_\odot$ all the way up to a maximum mass of $6\cdot 10^5 M_\odot$.
On the outer radius of the dyadosphere the electromagnetic field strength
becomes
\be
{\cal E}_{crit}=\frac{Q}{r_{ds}^2}
\hspace{1ex}.
\ee
It was shown that the electron-positron pair creation processes occur over the entire dyadosphere.
Thus an energy extraction process takes place due to the vacuum fluctuations in this region.
It was claimed that this energy source might lead to a natural explanation for the gamma ray
bursts as well as for the ultra high energy cosmic rays.
\par\noindent
In the presence of a strong electromagnetic field, as that in the dyadosphere, the velocity of light propagation
depends on the vacuum polarization states \cite{Adler:1971wn}. Drummond and Hathrell showed that the effect
of the vacuum polarization may lead to superluminal photon propagation \cite{Drummond:1979pp}.
Daniels and Shore who investigated the photon propagation around a charged black hole proved that
the effect of the one-loop vacuum polarization on photon propagation in the Reissner-Nordstr{\"o}m black hole
background makes the superluminal photon propagation possible \cite{Daniels:1993yi}.
Therefore, De Lorenci, Figueiredo, Fliche, and Novello were led to investigate the corrections for the
Reissner-Nordstr{\"o}m black hole metric due to the first contribution of the weak field limit of
the one-loop QED \cite{DeLorenci:2001bd}. They found that the metric element (\ref{metric1}) is written as
\be
f(r)=1-\frac{2 M}{r} + \frac{Q^2}{r^2}-\frac{\sigma Q^4}{5 r^5}
\hspace{1ex}.
\label{metric2}
\ee
The last term in (\ref{metric2}) expresses the contribution coming from the one-loop QED in the first
order of approximation. This term is of the same order of magnitude as the classical Reissner-Nordstr{\"o}m
charge term, i.e. the second term in (\ref{metric2}), as shown in \cite{DeLorenci:2001bd}.
\par\noindent
It is evident that the Reissner-Nordstr{\"o}m case arises from the aforementioned
metric element for the limit case $\sigma=0$.
%%%%%%%%%%%%%%%%%%%%%%%%%%%%%%%%%%%%%%%%%%%%%%%%%%%%%%%%%%%%%%%%%%%%%%%%%%%%%%%%%%%%%%%%%%%%%%%%%
%%%%%%%%%%%%%%%%%%%%%%%%%%%%%%%%%%%%%%%%%%%%%%%%%%%%%%%%%%%%%%%%%%%%%%%%%%%%%%%%%%%%%%%%%%%%%%%%%
%%%%%%%%%%%%%%%%%%%%%%%%%%%%%%%%%%%%%%%%%%%%%%%%%%%%%%%%%%%%%%%%%%%%%%%%%%%%%%%%%%%%%%%%%%%%%%%%%
\section{What has been done till now}
In 2003, Xulu evaluated the energy distribution in the dyadosphere for the modified Reissner-Nordstr{\"o}m
black hole metric \cite{Xulu:2003ma}. For this computation, the energy-momentum prescriptions of Einstein, Landau-Lifshitz,
Papapetrou, and Weinberg were employed. All four prescriptions gave the same and acceptable energy
distribution which is of the form
\be
E_{Einstein}=E_{LL}=E_{Pap}=E_{Wein}=M-\frac{Q^2}{2r}+\frac{\sigma Q^4}{10r^5}
\label{effmass1}
\hspace{1ex}.
\label{xulu}
\ee
It is clear that in the dyadosphere (where $r$ is small) the last term in (\ref{xulu}) plays an important role.
\par\noindent
By setting $\sigma=0$, one gets the energy distribution for the Reissner-Nordstr{\"o}m
black hole metric as evaluated by Virbhadra \cite{shwetket}, i.e.
\be
E_{R-N}=M-\frac{Q^2}{2r}
\hspace{1ex},
\label{rn1}
\ee
using the energy-momentum prescriptions of Einstein and Landau-Lifshitz.
%%%%%%%%%%%%%%%%%%%%%%%%%%%%%%%%%%%%%%%%%%%%%%%%%%%%%%%%%%%%%%%%%%%%%%%%%%%%%%%%%%%%%%%%%%%%%%%%%
%%%%%%%%%%%%%%%%%%%%%%%%%%%%%%%%%%%%%%%%%%%%%%%%%%%%%%%%%%%%%%%%%%%%%%%%%%%%%%%%%%%%%%%%%%%%%%%%%
%%%%%%%%%%%%%%%%%%%%%%%%%%%%%%%%%%%%%%%%%%%%%%%%%%%%%%%%%%%%%%%%%%%%%%%%%%%%%%%%%%%%%%%%%%%%%%%%%
\section{Energy distribution in M{\o}ller's Prescription}
The aim of this section is to evaluate the energy and momentum distributions associated with the
modified Reissner-Nordstr{\"o}m black hole metric, i.e. (\ref{metric}) and (\ref{metric2}),
using the M{\o}ller's energy-momentum complex. We first have to evaluate
the superpotentials in the context of M{\o}ller's prescription.
There are eight nonzero superpotentials
\bea
\xi^{1\,2}_{1}&=&-\xi^{2\,1}_{1}=2(M - \frac{Q^2}{r} +\frac{3\sigma Q^4}{5 r^5})\sin\theta\nn\\
\xi^{2\,3}_{3}&=&-\xi^{3\,2}_{3}=2(2M - \frac{Q^2}{r} +\frac{\sigma Q^4}{5 r^5}-r)\sin\theta\label{super}\\
\xi^{2\,4}_{4}&=&-\xi^{4\,2}_{4}=2(2M - \frac{Q^2}{r} +\frac{\sigma Q^4}{5 r^5}-r)\sin\theta\nn\\
\xi^{3\,4}_{4}&=&-\xi^{4\,3}_{4}=-2 \cos\theta\nn
\hspace{1ex}.
\eea

\par\noindent
By substituting the M{\o}ller's superpotentials, as given by (\ref{super}), into equation (\ref{mtheta}),
one gets the energy density distribution
\be
\mathcal{J}^{0}_{0}=\frac{Q^4 \left(r^4 -3\sigma Q^2\right)}{4\pi r^6}\sin\theta
\label{energyden}
\ee
while the momentum density distributions take the form
\bea
\mathcal{J}^{0}_{1}&=&0\label{momden1}\\
\mathcal{J}^{0}_{2}&=&0\label{momden2}\\
\mathcal{J}^{0}_{3}&=&0\label{momden3}\hspace{1ex}.
\eea

\par\noindent
Therefore, if we substitute equation (\ref{energyden}) into equation (\ref{menergy}),
we get the energy distribution of the modified Reissner-Nordstr{\"o}m black hole
that is contained in a ``sphere'' of radius $r$
\be
E(r)=M-\frac{Q^2}{r}+\frac{3\sigma Q^4}{5r^5}
\label{effmass2}
\ee
which is also the energy (mass) of the gravitational field that a neutral particle
experiences at a finite distance $r$. Thus, the energy given
by equation (\ref{effmass2}) is in addition called effective gravitational mass ($E=M_{eff}$)
of the spacetime under study.
\par\noindent
Additionally, if we replace equations (\ref{momden1}-\hspace{-0.1ex}
\ref{momden3}) into equation (\ref{mmomentum})
we get the momentum components which are given by
\bea
P_{1}=P_{2}=P_{3}=0\hspace{1ex}.
\eea
It is obvious that imposing $\sigma=0$ in expression (\ref{effmass2}) one gets the energy
distribution for the Reissner-Nordstr{\"o}m
black hole metric as evaluated by Virbhadra \cite{shwetket1}, i.e.
\be
E_{R-N}=M-\frac{Q^2}{r}
\hspace{1ex},
\label{rn2}
\ee
using the energy-momentum prescription of M{\o}ller.
\par\noindent
A couple of comments are in order. Firstly, a neutral test particle experiences at a finite distance $r$
the gravitational field of the effective gravitational mass described by expression (\ref{effmass2}).
Secondly, the energy-momentum complex of M{\o}ller as formulated here for the modified Reissner-Nordstr{\"o}m
black hole due to the first contribution of the weak field limit of
one-loop QED satisfies the local conservation laws
\be
\frac{\partial \mathcal{J}^{\mu}_{\nu}}{\partial x^{\mu}}=0
\hspace{1ex}.
\ee
Thirdly, it is easily seen that asymptotically  the effective gravitational mass of
the modified Reissner-Nordstr{\"o}m black hole takes the value of the Schwarzschild mass
\be
E(r\rightarrow\infty)=M
\hspace{1ex}.
\ee
Fourthly, comparing order by order the expressions (\ref{effmass1}) and (\ref{effmass2}) it is evident that
the coefficients of the r-terms are not equal. This discrepancy was identified some years ago when expressions
for the Reissner-Nordstr{\"o}m black hole, i.e. (\ref{rn1}) and (\ref{rn2}) were compared (see \cite{shwetket1}).
Therefore, we hypothesize that there is a connection between the coefficients of the expression for the
energy (when the energy-momentum complex of Einstein is employed) of the form
\be
E(r)= \sum_{n=0}^{+\infty} \alpha^{(Einstein)}_{n}  r^{-n}
\label{coeff1}
\ee
and those of the expression for the energy (when the energy-momentum of M{\o}ller is employed) of the form
\be
E(r)= \sum_{n=0}^{+\infty} \alpha^{(M{\o}ller)}_{n} r^{-n}
\label{coeff2}
\hspace{1ex}.
\ee
The relation that materializes this connection between the aforementioned coefficients is given by
\be
\alpha^{(Einstein)}_{n}=\frac{1}{n+1}\,\alpha^{(M{\o}ller)}_{n}
\label{connect}
\hspace{1ex}.
\ee
By comparing the result presented in this work, i.e. expression (\ref{effmass2}),
with the corresponding one that already exists in the literature, i.e. expression (\ref{effmass1}), is
easily seen that relation (\ref{connect}) holds true. In addition, the hypothesis is true for the expressions
associated with the Reissner-Nordstr{\"o}m black hole, i.e. (\ref{rn1}) and (\ref{rn2}).
Furthermore, in support of this hypothesis we provide one more example that already exists in the literature.
It was shown by Radinschi \cite{irina1} that the energy distribution of a charged regular black hole when
the energy-momentum complex of Einstein is employed, is given by
\be
E(r)=M-\frac{Q^2}{2r}+\frac{Q^6}{24 M^2 r^3}- \frac{Q^{10}}{240 M^4 r^5} +O\left(\frac{1}{r^6}\right)
\ee
while the corresponding energy distribution when the energy-momentum complex of M{\o}ller is employed, is given by
\be
E(r)=M-\frac{Q^2}{r}+\frac{Q^6}{6 M^2 r^3}- \frac{Q^{10}}{40 M^4 r^5} +O\left(\frac{1}{r^6}\right)
\hspace{1ex}.
\ee

%%%%%%%%%%%%%%%%%%%%%%%%%%%%%%%%%%%%%%%%%%%%%%%%%%%%%%%%%%%%%%%%%%%%%%%%%%%%%%%%%%%%%%%%%%%%%%%%%
%%%%%%%%%%%%%%%%%%%%%%%%%%%%%%%%%%%%%%%%%%%%%%%%%%%%%%%%%%%%%%%%%%%%%%%%%%%%%%%%%%%%%%%%%%%%%%%%%
%%%%%%%%%%%%%%%%%%%%%%%%%%%%%%%%%%%%%%%%%%%%%%%%%%%%%%%%%%%%%%%%%%%%%%%%%%%%%%%%%%%%%%%%%%%%%%%%%
\section{Conclusions}
In this work, we briefly presented the concept of the dyadosphere. This is a region outside of the
event horizon of the Reissner-Nordstr\"om black hole. In this region the electromagnetic field
strength is overcritical and thus electron-positron pair creation processes take place.
Therefore, the effects of the vacuum fluctuations in the dyadosphere modify the metric that describes the
Reissner-Nordstr\"om black hole. We explicitly calculate the energy and momentum density distributions associated
with the modified Reissner-Nordstr\"om black hole metric. The M{\o}ller's
energy-momentum complex is employed for this computation. The energy distribution derived here can be regarded as the
effective gravitational mass experienced by a neutral test particle placed in the spacetime under consideration.
In addition, setting the parameter $\sigma$ equals to zero, we derive the energy distribution
of the Reissner-Nordstr\"om black hole.
Furthermore, a comparison and a consequent
connection between our results that those already existing in the
literature is provided. We hypothesize that when the energy distribution is of specific form
(see expressions (\ref{coeff1}) and (\ref{coeff2})) there is a relation that connects the coefficients
in the Einstein's prescription with those in the M{\o}ller's prescription. This hypothesis is supported by the
expressions for the energy of the Reissner-Nordstr\"om black hole as well as for the energy of a charged regular black hole.
However, further investigation and verification is needed. We hope to return to this issue in a future work.
%%%%%%%%%%%%%%%%%%%%%%%%%%%%%%%%%%%%%%%%%%%%%%%%%%%%%%%%%%%%%%%%%%%%%%%%%%%%%%%%%%%%%%%%%%%%%%%%%%%%%%
%%%%%%%%%%%%%%%%%%%%%%%%%%%%%%%%%%%%%%%%%%%%%%%%%%%%%%%%%%%%%%%%%%%%%%%%%%%%%%%%%%%%%%%%%%%%%%%%%%%%%%
%%%%%%%%%%%%%%%%%%%%%%%%%%%%%%%%%%%%%%%%%%%%%%%%%%%%%%%%%%%%%%%%%%%%%%%%%%%%%%%%%%%%%%%%%%%%%%%%%%%%%%
\section*{Acknowledgements}
Research for ECV is supported by EPEAEK II
in the framework of the grant PYTHAGORAS II - University Research Groups Support
(co-financed $75\%$ by EU funds and $25\%$ by National funds).
%%%%%%%%%%%%%%%%%%%%%%%%%%%%%%%%%%%%%%%%%%%%%%%%%%%%%%%%%%%%%%%%%%%%%%%%%%%%%%%%%%%%%%%%%%%%%%%%%%%%%%
%%%%%%%%%%%%%%%%%%%%%%%%%%%%%%%%%%%%%%%%%%%%%%%%%%%%%%%%%%%%%%%%%%%%%%%%%%%%%%%%%%%%%%%%%%%%%%%%%%%%%%
%%%%%%%%%%%%%%%%%%%%%%%%%%%%%%% BIBLIOGRAPHY %%%%%%%%%%%%%%%%%%%%%%%%%%%%%%%%%%%%%%%%%%%%%%%%%%%%%%%%%

%%%%%%%%%%%%%%%%%%%%%%%%%%%%%%%%%%%%%%%%%%%%%%5
\end{document}